\documentclass[12pt]{article}
\usepackage{amsfonts}
\usepackage{amssymb}
\usepackage{graphics,amsmath}

\usepackage[utf8]{inputenc}
\usepackage[english]{babel}
\usepackage{a4wide}
\usepackage{amsfonts}
\usepackage{amssymb}
\usepackage{graphics,amsmath}
\usepackage{graphicx}
\usepackage{cancel}
\usepackage{cite}
\usepackage{mathrsfs}
\usepackage[usenames,dvipsnames,svgnames,table]{xcolor}
\usepackage{slashed}
\usepackage{etoolbox}
\usepackage{physics}
\usepackage[T1]{fontenc}

\def\hybrid{\topmargin -20pt    \oddsidemargin 0pt
        \headheight 0pt \headsep 0pt
        \textwidth 6.35in       
        \textheight 9.25in       
        \marginparwidth .875in
        \parskip 5pt plus 1pt   \jot = 1.5ex}

\hybrid

\def\baselinestretch{1.2}

\catcode`\@=11

\def\marginnote#1{}
\newcount\hour
\newcount\minute
\newtoks\amorpm
\hour=\time\divide\hour by60
\minute=\time{\multiply\hour by60 \global\advance\minute by-\hour}
\edef\standardtime{{\ifnum\hour<12 \global\amorpm={am}%
        \else\global\amorpm={pm}\advance\hour by-12 \fi
        \ifnum\hour=0 \hour=12 \fi
        \number\hour:\ifnum\minute<10 0\fi\number\minute\the\amorpm}}
\edef\militarytime{\number\hour:\ifnum\minute<10 0\fi\number\minute}

\def\draftlabel#1{{\@bsphack\if@filesw {\let\thepage\relax
   \xdef\@gtempa{\write\@auxout{\string
      \newlabel{#1}{{\@currentlabel}{\thepage}}}}}\@gtempa
   \if@nobreak \ifvmode\nobreak\fi\fi\fi\@esphack}
        \gdef\@eqnlabel{#1}}
\def\@eqnlabel{}
\def\@vacuum{}
\def\draftmarginnote#1{\marginpar{\raggedright\scriptsize\tt#1}}

\def\draft{\oddsidemargin -.5truein
        \def\@oddfoot{\sl preliminary draft \hfil
        \rm\thepage\hfil\sl\today\quad\militarytime}
        \let\@evenfoot\@oddfoot \overfullrule 3pt
        \let\label=\draftlabel
        \let\marginnote=\draftmarginnote
   \def\@eqnnum{(\theequation)\rlap{\kern\marginparsep\tt\@eqnlabel}%
\global\let\@eqnlabel\@vacuum}  }

\def\preprint{\twocolumn\sloppy\flushbottom\parindent 2em
        \leftmargini 2em\leftmarginv .5em\leftmarginvi .5em
        \oddsidemargin -.5in    \evensidemargin -.5in
        \columnsep .4in \footheight 0pt
        \textwidth 10.in        \topmargin  -.4in
        \headheight 12pt \topskip .4in
        \textheight 6.9in \footskip 0pt
        \def\@oddhead{\thepage\hfil\addtocounter{page}{1}\thepage}
        \let\@evenhead\@oddhead \def\@oddfoot{} \def\@evenfoot{} }

\def\numberbysection{\@addtoreset{equation}{section}
        \def\theequation{\thesection.\arabic{equation}}}

\def\underline#1{\relax\ifmmode\@@underline#1\else
        $\@@underline{\hbox{#1}}$\relax\fi}

\def\titlepage{\@restonecolfalse\if@twocolumn\@restonecoltrue\onecolumn
     \else \newpage \fi \thispagestyle{empty}\c@page\z@
        \def\thefootnote{\fnsymbol{footnote}} }

\def\endtitlepage{\if@restonecol\twocolumn \else \newpage \fi
        \def\thefootnote{\arabic{footnote}}
        \setcounter{footnote}{0}}  

\catcode`@=12
\relax

\def\figcap{\section*{Figure Captions\markboth
        {FIGURECAPTIONS}{FIGURECAPTIONS}}\list
        {Figure \arabic{enumi}:\hfill}{\settowidth\labelwidth{Figure
999:}
        \leftmargin\labelwidth
        \advance\leftmargin\labelsep\usecounter{enumi}}}
 \relax
\def\tablecap{\section*{Table Captions\markboth
        {TABLECAPTIONS}{TABLECAPTIONS}}\list
        {Table \arabic{enumi}:\hfill}{\settowidth\labelwidth{Table
999:}
        \leftmargin\labelwidth
        \advance\leftmargin\labelsep\usecounter{enumi}}}
 \relax
\def\reflist{\section*{References\markboth
        {REFLIST}{REFLIST}}\list
        {[\arabic{enumi}]\hfill}{\settowidth\labelwidth{[999]}
        \leftmargin\labelwidth
        \advance\leftmargin\labelsep\usecounter{enumi}}}
 \relax
\makeatletter
\newcounter{pubctr}
\def\publist{\@ifnextchar[{\@publist}{\@@publist}}
\def\@publist[#1]{\list
        {[\arabic{pubctr}]\hfill}{\settowidth\labelwidth{[999]}
        \leftmargin\labelwidth
        \advance\leftmargin\labelsep
        \@nmbrlisttrue\def\@listctr{pubctr}
        \setcounter{pubctr}{#1}\addtocounter{pubctr}{-1}}}
\def\@@publist{\list
        {[\arabic{pubctr}]\hfill}{\settowidth\labelwidth{[999]}
        \leftmargin\labelwidth
        \advance\leftmargin\labelsep
        \@nmbrlisttrue\def\@listctr{pubctr}}}
 \relax
\makeatother
\newskip\humongous \humongous=0pt plus 1000pt minus 1000pt

\newif\ifdtup

\relax

\def\be{\begin{equation}}
\def\ee{\end{equation}}
\def\ba{\begin{eqnarray}}
\def\ea{\end{eqnarray}}

\def\no{\noindent}

\def\IR{\relax{\rm I\kern-.18em R}}
\def\II{\relax{\rm 1\kern-.35em1}}

\def\IR{\relax{\rm I\kern-.18em R}}
\def\inv{^{\raise.15ex\hbox{${\scriptscriptstyle -}$}\kern-.05em 1}}

\DeclareMathOperator{\e}{e}

\begin{document}

\begin{titlepage}
\begin{center}

\vskip .5in

{\LARGE The $SU(2)$ Wess-Zumino-Witten spin chain \\ sigma model}
\vskip 0.4in

{\bf Rafael Hern\'andez},  \phantom{x} {\bf Juan Miguel Nieto} \phantom{x} and \phantom{x} {\bf Roberto Ruiz} 
\vskip 0.1in

Departamento de F\'{\i}sica Te\'orica and IPARCOS \\
Universidad Complutense de Madrid \\
$28040$ Madrid, Spain \\
{\footnotesize{\tt rafael.hernandez@fis.ucm.es, juanieto@ucm.es, roruiz@ucm.es}}

\end{center}

\vskip .4in

\centerline{\bf Abstract}
\vskip .1in
\no
\noindent Classical strings propagating in $AdS_{3}\times S^{3}\times T^{4}$ supported with Neveu-Schwarz-Neveu-Schwarz flux are described by a Wess-Zumino-Witten model. 
In this note, we study the emergence of their semiclassical $SU(2)$ spectrally flowed sectors as the Landau-Lifshitz limit of the underlying quantum spin chain. 
We consider the propagator in the coherent state picture, and find that the time interval is discretized proportionally to the lattice spacing. In the Landau-Lifshitz limit, 
where both time and space become continuous, we derive a path integral representation of the propagator for each spectrally flowed sector. We prove that the arbitrariness 
of the global phase of coherent states is mapped to the gauge freedom of the $B$-field in the classical action. We show that higher order corrections in the Landau-Lifshitz limit 
are suppressed by inverse powers of the 't Hooft coupling.  
\vskip .4in
\noindent

\end{titlepage}

\vfill
\eject

\def\baselinestretch{1.2}
%


\baselineskip 20pt


\noindent The $SL(2,\mathbb{R})$ Wess-Zumino-Witten (WZW) model constitutes a paradigm among exactly solvable realizations of string theory. 
It represents bosonic strings propagating in ${AdS}_{3}$ supported with Neveu-Schwarz-Neveu-Schwarz (NS-NS) three-form flux, and its spectrum is attainable 
through the study of the representations of the current algebra upon the worldsheet~\cite{MO}. 
The use of integrability in the AdS/CFT correspondence has renewed the interest in this model, since it rises in type IIB superstring theory realized 
on the integrable $AdS_{3} \times S^{3} \times T^{4}$ background with mixed three-form fluxes \cite{Cagnazzo}. Specifically, it corresponds to the pure NS-NS flux limit of the anti-de Sitter 
component of the bosonic truncation of the latter. In fact, a proposal for obtaining the spectrum of quantum strings on the whole background in that limit has been recently put forward on the basis of 
an integrable spin chain representation \cite{Baggio,Dei}, both matching and generalizing the previously existing results \cite{MO,Giribet}. 
This picture poses the problem of the emergence of the classical integrable structure, reached as the pure NS-NS flux limit of the mixed flux classical setting, 
arising from the worldsheet spin chain description. An appealing answer in this respect comes from the application of the Landau-Lifshitz limit to the spin chain. This method proved the equivalence 
between various sectors of four-dimensional $\mathcal{N}=4$ Yang-Mills theory and type IIB strings in the $AdS_{5} \times S^{5}$ background 
\cite{Kruczenski, Ryzhov, Rafa, nadgoB, Tseytlin, Bellucci, Esperanza, Park, Bogdan, BCM, MTT, MTT2, Tirziu, BC, Junior}. As reviewed in \cite{Treview}, 
the procedure starts with the expectation value of the Hamiltonian of the spin chain 
in a general coherent state. A path integral over periodic coherent states is then derived by means of the usual discretization of the time interval. 
The matching with the effective action of the slow degrees of freedom coming from the sigma model is performed 
when a continuum limit for the spin chain sites is taken in the action in the path integral. Such a matching can be traced to the rearrangement of the perturbative expansions of both sides of the duality 
in an effective parameter which can be made small from the two points of view. The method has the advantage of showing the equivalence between 
integrable structures avoiding the reference to any particular solutions, since the identification of the two actions entails the correspondence between their respective conserved charges.

In this note we will follow this path to retrieve the classical WZW model for strings on $\mathbb{R}\times S^{3}$  
with pure NS-NS flux from the the $SU(2)$ sector of the underlying spin chain  \cite{Hoare}. We will exploit the fact that the Landau-Lifshitz limit can be applied to any spin chain 
whose Hamiltonian is known to obtain an effective sigma model action for the worldsheet spin chain. In particular, we will derive a semiclassical path integral representation 
for the scalar product of two coherent states at different instants in every spectrally flowed sector of the theory. 
The partition of the time interval will reveal that time steps must be discretized in terms of the distance between spin chain sites. 
We will then apply the Landau-Lifshitz limit, which consists in simultaneous time and space continuum limits of the discrete expression.
This feature is in sharp contrast to the way it is displayed in the AdS$_{5}$/CFT$_{4}$ correspondence \cite{Kruczenski}, where the time interval is not discretized but partitioned, 
and the Landau-Lifshitz limit is introduced as a space continuum approximation in an exact path integral representation. We will find both the path integral representation and the condition 
under which it belongs to a specific spectrally flowed sector. We will show that the continuous counterpart of global phases of coherent states realizes the classical gauge field of the $B$-field.
We will further argue that subleading terms in the Landau-Lifshitz limit should emerge as quantum corrections. 

We will start discussing briefly the worldsheet spin chain which realizes the $SU(2)$ sector of the WZW model with pure NS-NS flux. 
The Hamiltonian under consideration is given by the shortening condition \cite{Lloyd}
\be
\label{Condition}
H^2=\left(\frac{k}{2\pi} P+M \right)^2 \ ,   
\ee
where $k=\sqrt{\lambda}$ is the level of the WZW model, with $\lambda$ the 't Hooft coupling, $P$ is the momentum operator and $M$ is an operator 
accounting for representation-dependent shifts of the dispersion relation. The level $k$ bounds the number of sites $L$ of the $w$-th spectrally flowed sector of the spin chain through \cite{MO,Dei}
\be
\label{Level}
kw+1\le L\le k(w+1)-1 \ ,    
\ee
where $w$ is a positive integer that parameterizes the spectral flow. Regarding the eigenvalues $\mu$ of $M$ for states of the $SU(2)$ spin chain, $\mu=0$ corresponds to the vacuum state, as a consequence of the BPS condition, 
while $\mu^2=1$ for single-magnon states \cite{Dei}. 
When $\mu=1$ the state transforms in a representation of the left-handed algebra of $SU(2)$, whereas it transforms under the right-handed one if $\mu=-1$. 
The action of $M$ on composite-magnon states follows from these relations. We are thus able to construct a $SU(2)$ spin chain of a given handedness by restricting the eigenvalue $\mu$ 
of the single-magnon states. The connection between $\mu$ and the angles in the Hopf fibration of the sphere, required for the construction below, can be elucidated as follows. 
Consider the metric of $S^{3}$ expressed as
\be
\textnormal{d} s^{2}=\frac{\left(4-{y_{1}^2-y_{2}^2}\right)^2\textnormal{d}\varphi_{1}^2+16\left(\textnormal{d}y_{1}^2+\textnormal{d}y_{2}^2\right)}{\left(4+{y_{1}^2+y_{2}^2}\right)^2} \ ,
\ee
where $\varphi_{1}$ is the angle of the Cartan torus in $S^{3}$ with respect to which the light-cone gauge is imposed, and $y_{i}$ are the coordinates in the transverse directions. 
The ranges of these variables are $\varphi_{1}\in[0,2\pi)$, $y_{i} \in [-2,2]$. When the canonical light-cone gauge quantization scheme is applied, the left-handed and right-handed choices of $\mu$ correspond, 
respectively, to excitations in the $Y=-y_{1}-iy_{2}$ and $\bar{Y}=-y_{1}+iy_{2}$ complex directions \cite{Lloyd}. The transformation to the coordinates in 
the Hopf fibering coordinates of the sphere, where the metric reads
\be
\label{Hopf}
\textnormal{d} s^{2} = \textnormal{d} \theta^2+\sin^2\theta\,\textnormal{d}\varphi_{1}^2+\cos^2\theta\,\textnormal{d}\varphi_{2}^2 \ , 
\ee
with $\theta\in[0,\pi/2]$, $\varphi_{i} \in[0,2\pi)$, is given by
\be
y_{1} = \frac{2\cos\theta}{1+\sin\theta}\cos\varphi_{2} \ , \quad y_{2} = \frac{2\cos\theta}{1+\sin\theta}\sin\varphi_{2} \ .
\ee
Therefore, the orientation of $\varphi_{2}$ accounts for the handedness of the $SU(2)$ representation. 

In order to map the spin chain to a sigma model we need to introduce a continuum set of variables at each site of the chain, as reviewed for instance in \cite{AF}. 
An infinite set of states can be constructed by applying a $SU(2)$ rotation to an eigenstate of the Cartan generators. If we parameterize such rotation in terms of Euler angles, the one-site coherent state reads \cite{AF}
\be
\label{Vector}
\ket{\vec{n}}=\e^{i\chi}\left(\e^{i\varphi}\cos\vartheta\ket{1}+\e^{-i\varphi}\sin\vartheta\ket{2}\right) \ ,
\ee
where $\ket{1}$ and $\ket{2}$ span the fundamental representation of $SU(2)$, and the Euler angles are related with those in (\ref{Hopf}) as
\be
\label{Euler}
{\vartheta}=\frac{\pi}{2}-\theta \ , \quad \varphi=\frac{\varphi_{1}+\varphi_{2}}{2} \ , \quad {\chi}=\frac{\varphi_{1}-\varphi_{2}}{2} \ .
\ee
Coherent states form an overcomplete basis with the resolution of the identity 
\be
\label{Spectral}
\int\textnormal{d}\mu[\vec{n}]\ket{\vec{n}}\bra{\vec{n}} = \mathbb{I} \ ,        
\ee
where
\be
\label{Measure}
\textnormal{d}\mu[\vec{n}]=\frac{1}{\pi^2}\sin2\vartheta\, \textnormal{d}\vartheta\, \textnormal{d}\varphi\,\textnormal{d}\chi \ ,
\ee
is the Haar measure. The extension to coherent states in a spin chain with $L$ sites is obtained from the tensor product of $L$ one-site coherent states and the associated measure 
is therefore equal to the product of one-site measures. It should be noted that the angles $\chi$ appear as global phases of one-site coherent states. If they were removed
as in the case of $\mathcal{N}=4$ Yang-Mills, we would find the $SU(2)/U(1)$ Landau-Lifshitz model rather than the $SU(2)$ one in the subsequent derivation. 
Here we will keep them as they will be crucial in the comparison with the Landau-Lifshitz limit of the non-linear classical sigma model. We also note that the 
operation $\varphi_{2}\mapsto-\varphi_{2}$ that inverts the handedness of the $SU(2)$ representation is realized by the exchange of
$\varphi$ and $\chi$ at each site.

The propagator of the theory to which the path integral analysis is applied merges an initial coherent state of the spin chain at time $t_{0}=0$, $\ket{\vec{n}_{0,a}}$, 
and a final one at the instant $t_{N}=T$, $\ket{\vec{n}_{N,a}}$,~\footnote{We perform the derivation in Minkowskian time. 
The path integral expression that we would obtain in Euclidean time can be recovered from the one presented here by a Wick rotation.} 
\be
\label{Generating}
Z=\bra{\vec{n}_{N,a}}\exp\left({-iT H}\right)\ket{\vec{n}_{0,a}} \ ,
\ee
where $a=0,...,L-1$ labels the spin chain sites. We impose periodic boundary conditions $\vec{n}_{0,L}=\vec{n}_{0,0}$ and $\vec{n}_{N,L}=\vec{n}_{N,0}$ so that coherent states depict
closed strings and limits thereof in the classical sigma model limit. Moreover, we will choose the spin chain frame rather than the string frame, as it will prove 
best suited to the setting considered here \cite{Dei}.

Now we must recall that the Hamiltonian above is taken to be realized in the proper representation of the algebra. 
However, the shortening condition (\ref{Condition}) does not provide an explicit expression for the Hamiltonian 
in terms of the momentum operator, but for its square. Assuming the non-negativeness of the former, 
the sign of the linear relation between these operators cannot be determined without any reference to state vectors. 
We can circumvent this hindrance in (\ref{Generating}) by requiring $\ket{\vec{n}_{0,a}}$ (or $\ket{\vec{n}_{N,a}}$) 
to be an eigenstate both of the momentum operator and the Hamiltonian. Such requirement of course constrains the admissible values 
of $\vartheta_{0,a}$ and $\varphi_{0,a}$ (respectively $\vartheta_{N,a}$ and $\varphi_{N,a}$). Given, for instance, 
the relation between the energy and the eigenvalue $p$ of the momentum (in principle quantized in view 
of the aforementioned boundary conditions), 
\be
\label{Dispersion}
E = - \frac{k}{2\pi}p-\mu \ ,
\ee
where $\mu$ denotes the eigenvalue of $M$ for the state $\ket{\vec{n}_{0,a}}$, we are able to write 
\be
\label{Hamiltonian}
Z = \bra{\vec{n}_{N,a}}\exp\left({iT\left({k}/{2\pi}\right) P + i T M }\right)\ket{\vec{n}_{0,a}} \ .
\ee
We will now slice $[0,T]$ in $N$ subintervals $[t_{\alpha+1},t_{\alpha}]$ of equal length $\Delta t$, with $t_{\alpha}= \alpha T/N$ and $\Delta t=T/N$. If we introduce an insertion of the spectral decomposition 
of the identity between the endpoints of every pair of consecutive subintervals, we can express (\ref{Hamiltonian})  as
\be
\label{Partition}
Z = \int\textnormal{d}\mu_{1}\,...\,\textnormal{d}\mu_{N-1}\overset{N-1}{{\underset{\alpha=0}{\prod}}}\bra{\vec{n}_{\alpha+1,a}}\exp\left({i\Delta t\left({k}/{2\pi}\right) P + i \Delta t M}\right)\ket{\vec{n}_{\alpha,a}} \ ,     
\ee
where $\textnormal{d}\mu_{\alpha}$ is the measure of the $\alpha$-th insertion. At this point we must emphasize that the time dependence must be intrinsically discretized for the spin chain at issue. 
This follows from the expression of the operator appearing in (\ref{Partition}),
which involves the shift operator $U=\exp\left(i\epsilon P \right)$ raised to the power $k\Delta t/ 2 \pi\epsilon$, where~$\epsilon$ is the lattice spacing. 
In order for the action of the shift operator on the state space to be defined, the length of the time subinterval must then satisfy 
\be
\label{Gap}
\Delta t=\frac{2\pi\epsilon}{k} \ ,  
\ee
which implies the discretization of the time interval. Note that even if $\Delta t=2\pi m\epsilon/k$, with $m$ a positive integer, 
is also admissible, the choice $m=1$ is always achievable by diving the whole interval $[0,T]$ in enough subintervals. 
For concreteness we will consider the right-handed $SU(2)$ spin chain. The operator in (\ref{Partition}) is then realizable via \footnote{The choice of one or other vacuum determines classical actions which differ
in a term without derivatives whose sign depends on the handedness of the $SU(2)$ representation. This is however irrelevant in the Landau-Lifshitz limit. Besides, whatever the choice is, it is meant to represent the BPS vacuum 
of the R-R sector in the center of the Hodge diamond \cite{Dei}.}
\be
\label{Addition}
\textnormal{exp}\left(i\epsilon P + i \Delta t \, \overset{L-1}{\underset{a=0}{\bigotimes}}\left(\sigma_{3}^{a}- \mathbb{I}^{a}\right)/2\right) 
= U\textnormal{exp}\left(i\Delta t \, \overset{L-1}{\underset{a=0}{\bigotimes}}\left(\sigma_{3}^{a}-\mathbb{I}^{a}\right)/2\right) \ ,
\ee
where $\sigma^{a}_{3}$ is the diagonal Pauli matrix acting on the $a$-th site. If we choose this representation, we can write the propagator (\ref{Partition}) as
\be
\begin{split}
\label{Second}
Z &= \int\textnormal{d}\mu_{1}\,...\,\textnormal{d}\mu_{N-1}\overset{N-1}{{\underset{\alpha=0}{\prod}}}\,\overset{L-1}{{\underset{a=0}{\prod}}}\e^{-i\left(\chi_{\alpha+1,a-1}-\chi_{\alpha,a}\right)}\left[\e^{-i\left(\varphi_{\alpha+1,a-1}-\varphi_{\alpha,a}\right)}\cos\vartheta_{\alpha+1,a-1}\cos\vartheta_{\alpha,a}\right.\\
& + \left. \e^{i\left(\varphi_{\alpha+1,a-1}-\varphi_{\alpha,a}+\Delta t\right)}\sin\vartheta_{\alpha+1,a-1}\sin\vartheta_{\alpha,a}\right] \ .
\end{split}
\ee
The leading order action of the Landau-Lifshitz sigma model is obtained in the semiclassical limit of long wavelength, 
where the spin chain sites are taken to be parameterized continuously and the dependence of the angles on them is assumed to be analytic. In a conventional spin chain this limit consists in extracting the leading contribution inside the path integral 
after sending $\epsilon\rightarrow0$ and $L\rightarrow\infty$, with the length of the spin chain $R=L\epsilon$ fixed. Such path integral is constructed by applying to the propagator a prior continuum limit of the time coordinate
defined by $\Delta t\rightarrow 0$ and $N\rightarrow\infty$, while $T=N\Delta t$ kept fixed. However, space and time continuum limits are intertwined for the spin chain considered here. 
On the one hand, space and time step lengths are proportional as stated by (\ref{Gap}), that implies that $T=2\pi NR/kL$. Therefore, when both continuum limits are applied the condition $N/kL\sim\mathcal{O}(1)$ is needed. 
On the other hand, the number of sites $L$ of the chain is bounded by the WZW level $k$ in view of equation (\ref{Level}), and thus $L\rightarrow \infty$ already assumes the semiclassical limit $k\rightarrow \infty$, 
which implies that $\Delta t\rightarrow 0$. In fact, in order for the spin chain to remain within the $w$-th spectrally flowed sector, it must be satisfied that 
\be
\label{Espectral}
w\le L/k\le w+1 \ ,
\ee
when both $L,k\rightarrow\infty$. If we apply the Landau-Lifshitz limit to (\ref{Second}) taking these observations into account, the propagator becomes
\be
\begin{split}
\label{Propagator}
Z& = \int\textnormal{d}\mu_{1}\,...\,\textnormal{d}\mu_{N-1}\overset{N-1}{{\underset{\alpha=0}{\prod}}}\,\overset{L-1}{{\underset{a=0}{\prod}}}\,
\left[\vphantom{\mathcal{O}\left(\epsilon^2\right)}1-i\Delta t\left({\dot{\chi}_{\alpha,a}}+\cos 2\vartheta_{\alpha,a}\dot{\varphi}_{\alpha,a}-\sin^2\vartheta_{\alpha,a}\right)\right.\\
& + \left. i\epsilon(\chi_{\alpha,a}'+\cos 2\vartheta_{\alpha,a}\varphi_{\alpha,a}')+\mathcal{O}\left(\epsilon^2\right)\right] \ ,
\end{split}
\ee
where the dot denotes the derivative with respect to $t_{\alpha}$, while the prime is the derivative with respect to $x_a=a\epsilon$. 
The integrand in (\ref{Propagator}) may be regarded as the formal product of two Volterra continuous products when $\Delta t,\epsilon\rightarrow 0$ once the short distance cut-off $1/\Delta t$, 
or equivalently $1/\epsilon$, is introduced \cite{AF}. Therefore, in the continuum limit we can write
\begin{equation}
\label{Continuum}
Z=\int\left[\textnormal{d}\mu\right] \e^{iS} \ ,
\end{equation}
where $\left[\textnormal{d}\mu\right]$ denotes the path integral measure and
\begin{equation}
\label{Funcional}
\begin{split}
S =-\frac{1}{\epsilon}\int_{0}^T{\textnormal{d}t}\int_{0}^{R}\textnormal{d}x\left[\dot{\chi} -\frac{k}{2\pi} \chi' +\cos2\vartheta \left(\dot{\varphi} -\frac{k}{2\pi}\varphi' \right) - \sin^2\vartheta \right] \ . 
\end{split}
\end{equation}
The path integral extends over configurations subject to the periodicity condition $\vec{n}(t,x)=\vec{n}(t,x+R)$ that satisfy the boundary conditions $\vec{n}(0,x)=\vec{n}_{0}(x)$ and $\vec{n}(T,x)=\vec{n}_{N}(x)$, 
where $\vec{n}_{0}(x)$ and $\vec{n}_{N}(x)$ are, respectively, the continuous counterparts of $\vec{n}_{0,a}$ and $\vec{n}_{N,a}$. It is worth to emphasize that we could have proceeded by expressing equation (\ref{Hamiltonian}) as
\begin{equation}
Z = \int\textnormal{d}\mu_{1}\,...\,\textnormal{d}\mu_{N}\overset{N-1}{{\underset{\alpha=0}{\prod}}}\exp\left({-iTE}\right) \bra{ \vec{n}_{\alpha+1,a} }\ket{ \vec{n}_{\alpha,a} } \ , 
\end{equation}
which leads to the same action as (\ref{Funcional}) once the expression of its associated Hamiltonian is considered. Nevertheless, if we had taken directly this path, some steps that are necessary in the attainment 
of equation (\ref{Funcional}) would have been obscured.

The Landau-Lifshitz limit can also be applied at the level of the classical string action.
It is obtained as the perturbative expansion with respect to $\sqrt{\lambda}/L$ of the Lagrangian density in the statically gauge fixed T-dual 
sigma model \cite{Tseytlin}. In particular, the Wess-Zumino term involving first order time derivatives appears at zeroth order, whereas the $n$-th order contribution consists in terms with $n$ spatial derivatives. 
Such an expansion can be achieved in the action above by means of the change of variables $t\mapsto t/k$, which enables us to compare (\ref{Funcional}) with the result obtained from the 
classical non-linear sigma model. If we introduce this change of scale together with $x\mapsto x/2\pi$, the leading order action in the semiclassical limit reads 
\begin{equation}
\label{Landau}
S=-\frac{1}{2\pi\epsilon}\int_{0}^\infty\textnormal{d}t\int_{0}^{2\pi R}\textnormal{d}x\left[\dot{\chi}+\cos2\vartheta\,\dot{\varphi} - \left(\chi' + \cos2\vartheta\,\varphi' \right)\right] \ .
\end{equation}
If we choose now $R=1$, we can replace $\epsilon$ by $1/L$ in front of the action, hence obtaining the NS-NS limit of the Landau-Lifshitz model found in \cite{Hoare}. 
The field accounting for the $U(1)$ gauge freedom in the choice of the $B$-field is mapped to the variable $\chi$ that realizes the continuous counterpart of global phases of coherent states. 
Here its presence is required to match the classical result, as opposed to the analogous scenario in the AdS$_{5}$/CFT$_{4}$ correspondence, where the $B$-field plays no role. 
In order to fix the gauge arbitrariness we may require, for instance, finiteness of the classical action evaluated over dyonic giant magnon solutions, which leads to the constraint $\chi=\varphi$ \cite{Hoare}. 
This condition is imposed because such configurations represent, in the spin chain picture, a bound state of a macroscopic amount of magnons over the spin chain vacuum. It is worth to stress that such gauge 
fixed action is invariant under the operation interchanging $\varphi$ and $\chi$, and hence it describes  both the left- and right-handed $SU(2)$ Landau-Lifshitz models.

The action (\ref{Landau}) is the zeroth order in $\sqrt{\lambda}/L$ of the Landau-Lifshitz model. In principle, it should be corrected by subleading terms in the long wavelength limit \cite{Ryzhov,Tseytlin}. 
However, these corrections are suppressed by additional powers of $1/\sqrt{\lambda}$ for the model at issue. 
We can show this by considering the most general possible term in (\ref{Propagator}), involving $A$ time derivatives, $B$ spatial derivatives 
and none derivatives in a factor $\phi^{C}$ accounting for the contribution of $M$. The contribution to the action of this term, 
$ \Delta t^{A+C}\epsilon^{B}\dot{\phi}^{A}\phi'^{B}\phi^{C} \ $, after the change of variables $t\mapsto t/k$ and $x\mapsto x/2\pi$ is made (and $R=1$ is set), reads 
\be
\frac{\left(2\pi\right)^{A+B+C-1}}{k^{A+B+2C-1}}\left(\frac{k}{L}\right)^{A+B+C-1}\frac{L}{2\pi}\int_{0}^{\infty}\textnormal{d}t\int_{0}^{2\pi}\textnormal{d}x\,\dot{\phi}^{A}\phi'^{B}\phi^{C} \ .
\ee
Therefore terms with either $A>1$, $B>1$ or $C>0$ carry additional powers of $1/\sqrt{\lambda}$ besides the factor $L(\sqrt{\lambda}/L)^{A+B+C-1}$, 
and thus cannot be obtained from the Landau-Lifshitz expansion of the classical non-linear sigma model. \footnote{A slight extension of the analysis performed in \cite{Hoare} shows that such is indeed the case.}

There are some natural extensions to the derivation in this letter. The most immediate question concerns the emergence of the classical Landau-Lifshitz model 
from other sectors of the worldsheet spin chain, such as the left-handed and right-handed $SL(2,\mathbb{R})$ sectors, or those which include fermions. 


\vspace{12mm}

\centerline{\bf Acknowledgments}

\vspace{2mm}

\noindent
J. M. Nieto and R. Ruiz would like to thank the organizers of the program {\em YRISW 2019: A modern primer for 2D CFT} 
at the Erwin Schr\"odinger Institute in Vienna for support while this work was being completed. 
The work of R.~H. is supported by grant FPA2014-54154-P and by BSCH-UCM through grant GR3/14-A 910770. 
R.~Ruiz acknowledges the support of the Universidad Complutense de Madrid through the predoctoral grant CT42/18-CT43/18.


\end{document}